\documentclass[12pt]{article}
\usepackage{color,multicol}
\usepackage{amsmath}
\usepackage{amssymb}
\usepackage{epsfig}

\newcommand{\set}[1]{{\mathbb{#1}}}

\begin{document}

\title{Learning, investments and derivatives}
\date{August 12, 2010\footnote{Date of the original submission to RISK Magazine.}}
\author{Andrei N. Soklakov}
\maketitle


\begin{center}
\parbox{14cm}{
{\small
The recent crisis and the following flight to simplicity put most
derivative businesses around the world under considerable
pressure. We argue that the traditional modeling techniques
must be extended to include product design. We propose a quantitative
framework for creating products which meet the challenge of being
optimal from the investors point of view while remaining relatively
simple and transparent.
} }
\end{center}

\section*{Introduction}
In the last few decades, the world of financial derivatives experienced
a true explosion in the development of quantitative methods.
Remarkably, very little of that development went into supporting
the process of product innovation and structuring
-- the cornerstone of every derivative business! Compared to modeling,
which continues to attract most quantitative effort,
product innovation remains essentially a form of art. Without diminishing
the importance of accurate modeling, it is equally important to recognize that
payoff structure of a financial product is not just a source of modeling
challenges, -- it is a central part of a quantitative solution, -- a part
which needs its own quantitative support.

Innovation going badly has been among the main themes of the
recent crisis. The time has come for a serious look at quantitative ways
of making the process of innovation more rational and systematic. The purpose of this article
is to introduce some elementary arguments which may form the basis of a more
rational quantitative approach to structuring.

For the sake of clarity and generality we develop our key arguments from
first principles. Going back to the beginning, we see two basic reasons for trading derivatives.
First one is the demand for bespoke hedging solutions. The other
is leveraged investments. Without jumping to conclusions
as to which of these two reasons is more or less natural, it would be fair to say
that the recent crisis raised serious concerns about the use of derivatives for
investment purposes. For this reason we investigate structuring investments as
our first priority.

Investment decisions are driven by peoples' beliefs as much as
by economic facts. We therefore start by looking for a simple mathematical language
that treats facts and beliefs on an equal footing. We show how beliefs can be
converted into tradable payoff structures and, conversely, how to imply
investors beliefs from the payoff structures they use. This gives us tools
for tailoring new products as well as checking implicit assumptions behind existing positions.

Most people would agree that complex products can be dangerous. What is less
obvious is that simple products can also be pretty far from ideal.
We find that investment products can be very easily oversimplified to the point
when they can no longer reflect any rational view.
Paraphrasing the famous observation about physical theories, investments
should be made as simple as possible but not simpler.

Two fundamental features distinguish investment from speculation:
the use of in-depth research and attention to safety. The above caution
on oversimplification is related to the ability of products to accommodate
results of in-depth research. Thinking about safety, we show how one can
use the technology of derivatives to stabilize raw investment strategies.
This line of thinking leads us to a class of derivatives which, in terms
of their complexity, lie between vanillas and exotics.

\section*{Investments and probability distributions}
Consider a random variable representing an observable market parameter:
stock price, index value, interest rate, etc.
Everything we know about the variable can be summarized using probability distributions.
The ability of probability distributions to summarize the knowledge about
random variables does not depend on whether this knowledge is factually true
or not. This powerful feature of probability-based
descriptions left its mark on most modern fields of human knowledge:
from foundations of quantum mechanics to artificial intelligence.
It is certainly very important for describing the world of finance where
both hard data and investors' beliefs have material impact.

Interplay between reality and beliefs is at the heart of
every investment. We set the stage by introducing
distinct concepts of market-implied, investor-believed and actual
realized distributions -- all of which refer to the same market variable.
These distributions play three distinct roles which can be outlined as follows.
Market-implied distribution, $m()$, represents the market view.
The reader is almost certainly familiar with deriving such
distributions from the relevant market prices. The investor's view is given
by the believed distribution, $b()$. The form of this distribution is
entirely up to the investor and we do not make any assumptions regarding
$b()$. Realized distribution, $p()$, represents
the reality that both the market and the investor are trying to predict.
In practice, realized distribution can be very difficult to estimate leading
to the use of proxies such as realized mean and realized variance.
We use $m()$, $b()$ and $p()$ as
auxiliary concepts -- as thinking tools -- refining the definitions along the way
as required. We shall see, however, that these concepts have very real
and easily accessible (often visualizable) meaning.

In hindsight, it is not
surprising to find products which already use the
language of distributions in their intuition. Futures and variance swaps, for example,
compare the mean and the variance of the market-implied and realized distributions.
Exotics, such as barriers, suggest that there might be more to trading
distributions than using a few of their leading moments. This is indeed the case.
Imagine, for example, that you have strong opinions on both the mean and the variance
of some market variable. How would you allocate your money? Should you put
most of it on the mean or on the variance? Are you even sure that your opinion
is better expressed with two separate trades (one on the mean and the other
on the variance), or is there a structured product that is optimized to express
the combined view? We shall see that thinking in terms of probability
distributions makes such questions trivial and to a large degree redundant.

\section*{Growth-optimizing investor}
One can talk about optimality of investments only with respect to an
objective. As illustrated by Markowitz, different objectives
normally lead to different optimal investments. On the practical side
this means a high focus on examples. The aim of this section is to
introduce an example of a particular investor which is then used as
a reference point supporting the arguments presented in the subsequent sections.
The case of a general investor is presented in~\cite{Soklakov_Forthcoming}.

Imagine an investor who considers himself an expert on a certain market
variable. The investor is very confident in his expertise and is seeking
a buy-and-hold product with two characteristics: the product must make use
of the entire expert knowledge about the variable and offer the highest
expected rate of return. Even without formal analysis it is clear that
this growth-optimizing investor is very aggressive. With a possible
exception of near-crisis bubble periods, when the rate of returns becomes
a common obsession, most people would be more risk averse.

Let us start our analysis with the simplest limiting case. Imagine that the
investor is absolutely sure that the value of the market variable, $x$,
at certain future time will lie in a small interval $[K_1,K_2]$.
Graphically, this belief can be described by the probability
distribution on Fig.~1A. The optimal investment strategy expressing this
belief must offer nonzero reward if $x$ falls within the interval and
pay zero otherwise. For small enough $\delta=K_2-K_1$ the variation of the
payment within the interval can be neglected and the whole strategy can be
viewed as buying a digital call spread with strikes $K_1$ and $K_2$.

In a slightly more general case, the investor might believe that two
small intervals $[K_1,K_2]$ and $[L_1,L_2]$ contain the future
outcome for $x$. This belief can be summarized by a pair of probabilities
$p_K$ and $p_L=1-p_K$ as illustrated by the distribution on Fig.~1B. As before,
it is clear that the investor should focus on combinations of two binary
spreads. The only nontrivial question here is how to split the
investment between the two spreads in order to maximize investor's utility.

In general, one can imagine a fine set of binary spreads covering the domain
of any reasonable probability distribution (Fig.~1C.). The problem of optimal
investment strategy (maximum expected utility) can then be formulated as
the problem of finding proportions of the total investment which should
be allocated to each of the binary spreads.

At the mathematical level, the above discussion connects the optimal investment
problem to the classic analysis of horse racing scenarios: just think of
the binary spreads as \lq\lq horses", with only one of them \lq\lq winning" i.e. maturing
in the money. The mathematics of such scenarios has been around for a long
time and is very well understood. In what follows we use this
understanding without assuming any special expertise from the~reader.\footnote{
Given the sensitivity of the post-crisis time of this publication, it is
worth emphasizing that our references to the mathematics of horse racing
are purely pedagogical (as is, for example, the famous
Prisoner's Dilemma in decision theory). There is no doubt that the relevant
pieces of logic can be introduced without any reference to gambling
(or prisoners). Such references however export the clarity of
mathematical understanding to a very wide audience -- a feature worth
keeping if we are to achieve maximum transparency.}

\begin{center}
    \includegraphics[scale=0.38]{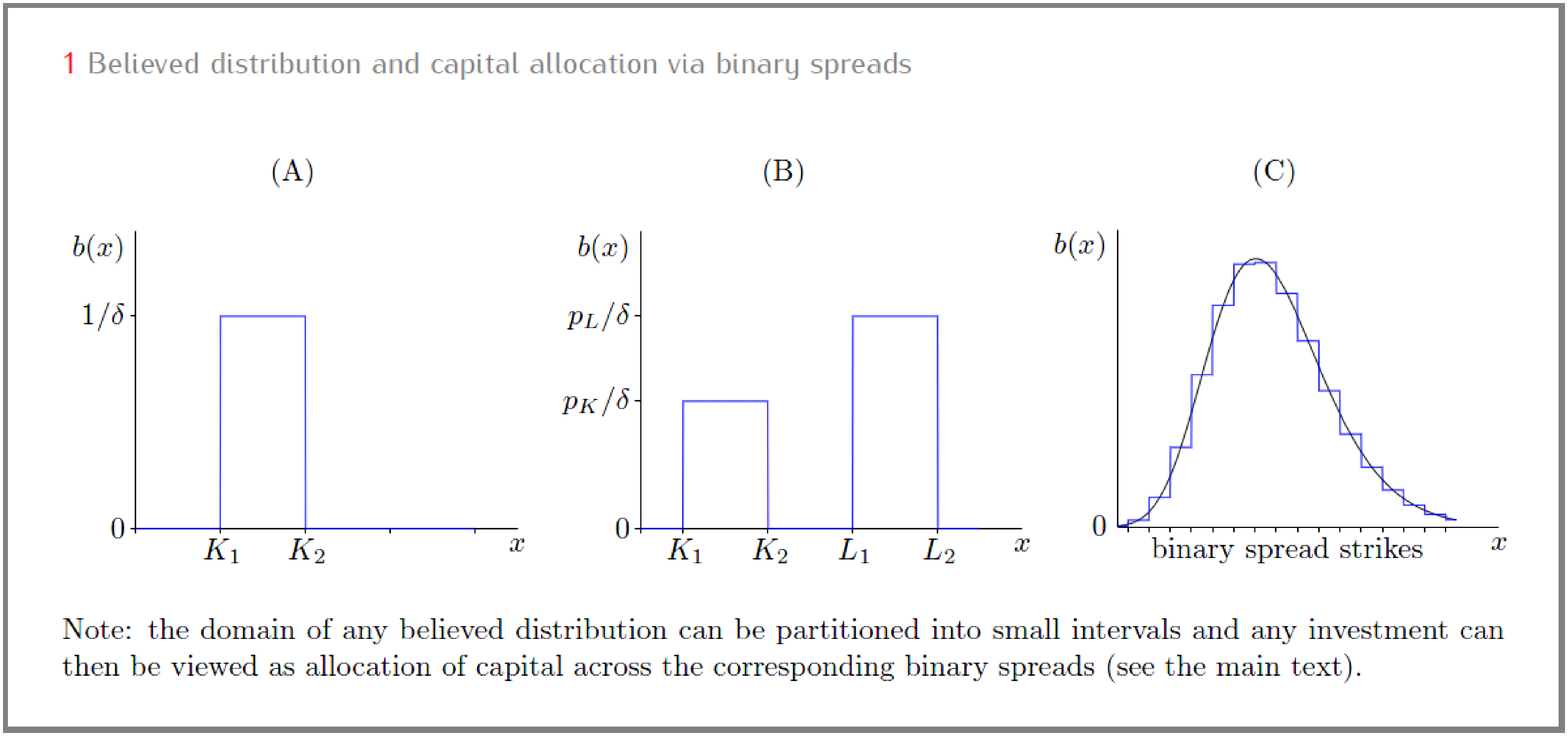}
\end{center}

Let $N$ be the number of binary spreads and let $\{R_i\}_{i=1}^N$ be the
returns on each of the spreads offered by the market. Furthermore,
let $\{\alpha_i\}_{i=1}^N$ be the
proportions in which the investor chooses to split their investment across
the spreads ($\sum_i\alpha_i=1$). If $b_i$ is the investor's believed
probability that the $i$th spread matures in the money, then for the
expected rate of return he would compute
\begin{equation}\label{Eq:EbRate}
\set{E}_b[{\rm rate}] = \sum_ib_i\ln(\alpha_i R_i)\,.
\end{equation}
Maximizing this over all possible $\{\alpha_i\}_{i=1}^N$ subject to the
constraint $\sum_i\alpha_i=1$ leads us to the classic result~\cite{Kelly_1956}
\begin{equation}\label{Eq:Alpha*}
\alpha_i^*\stackrel{\rm def}{=} {\rm optimal\ }\alpha_i=b_i\,.
\end{equation}
In other words, investment in proportion to the believed probabilities maximizes
the expected rate of return (all from the investor's point of view). In what
follows we will also need the expression for the actual expected return from
this strategy. This is
\begin{equation}
\set{E}_p[{\rm rate}] = \sum_ip_i\ln(\alpha^*_i R_i) = \sum_ip_i\ln(b_i R_i)\,,
\end{equation}
where $p_i$ are the actual realized probabilities which are unknown at the
time of investment. For simplicity, we start by looking at the \lq\lq fair odds'' case
when $\sum_i1/R_i=1$. This allows us to define the market-implied probability,
$m_i\stackrel{{\rm def}}{=}1/R_i$, that the $i$th spread
matures in the money and rewrite the above formula as
\begin{equation}\label{Eq:ERate}
\set{E}_p[{\rm rate}] = \sum_ip_i\ln\frac{b_i}{m_i}\,.
\end{equation}
In the above derivation of equations~(\ref{Eq:EbRate}-\ref{Eq:ERate}) we chose to make two simplifying assumptions: all of the investment
is allocated across the spreads (as opposed to keeping some of it as cash) and $\sum_i1/R_i=1$. We did this as a logical shortcut
which makes mathematics throughout this paper particularly transparent. In the Appendix we show what happens when these assumptions are removed:
some interesting subtleties emerge but the core logic remains intact. The reader is advised to continue with the main
text of the paper before reading the Appendix.

Speaking about assumptions, it is important to note that we do not use any pricing or econometric models.
For us, market-implied probabilities are nothing but a probabilistic way of recording market returns. It is just
like probabilistic interpretation of betting returns -- the odds. And just like people on the race track,
who discuss the odds offered by the bookie, we can speak about $m_i$ as long as the market offers the relevant prices.
This remains true even if the market is distorted by inefficiencies, including arbitrage and large bid-offer spreads.
Our definition of market-implied distribution is therefore valid even when a closely related concept of market-implied pricing measure
does not exist. Investor's belief probabilities, $b_i$, will be discussed in detail in the
next section. We will not need the realized distribution, $p_i$, until the
penultimate section of the paper.

\section*{The maximal payoff of learning}
In terms of probabilities, the process of learning is often described via Bayes' theorem
\begin{equation}\label{Eq:Bayes}
   P(x|{\rm research})=\frac{P({\rm research}|x)}{P({\rm research})}\,P(x)\,,
\end{equation}
which tells us how the prior knowledge $P(x)$ about random variable $x$
is updated to $P(x|{\rm research})$ on the account of learning from results
of additional research. This formula follows immediately from the properties
of conditional probabilities and is therefore valid for any consistent
learning process regardless of whether or not the actual implementation
makes use of~Eq.~(\ref{Eq:Bayes}).

In the previous section we computed the return to the growth-optimizing investor as
\begin{equation}\label{eq:fi}
    \alpha_i^*\, R_i=b_iR_i=\frac{b_i}{m_i}\,,
\end{equation}
where $i$ is the index of the binary spread which ends up maturing
in the money. Considering this index as a function of the underlying
market variable, $i=i(x)$, we can write the payoff~(\ref{eq:fi})
from a single run of the growth-optimizing strategy as
\begin{equation}\label{eq:f_from_b}
f(x)=\frac{b(x)}{m(x)}\,,
\end{equation}
where $b(x)=b_{i(x)}/\Delta x$ and $m(x)=m_{i(x)}/\Delta x$ are the believed
and the market-implied probability densities. Rearranging the terms, we obtain
\begin{equation}\label{eq:b_from_f}
b(x)=f(x)\,m(x)\,.
\end{equation}
Comparison of this formula to Eq.~(\ref{Eq:Bayes}) gives us the fundamental
connection between the payoff structure for the optimal growth strategy and
the underlying learning process. Indeed, any rational investor starts
by looking at market prices. Mathematically, this means that the investor is
taking the market-implied view as the prior, $P(x)=m(x)$.
Any logical alternative to this choice of prior would imply mistaking market information.
The investor then updates the prior by performing, purchasing or assuming additional
research. This step may or may not use Bayes' theorem in an explicit way.
However, if the investor is capable of writing the updated view as a probability distribution,
it would have to be the posterior belief $b(x)\stackrel{\rm def}{=}P(x|{\rm research})$.
This leaves us no choice but to recognize Eq.~(\ref{eq:b_from_f}) as a special case of
Eq.~(\ref{Eq:Bayes}). The general case of an arbitrary investor is subject to the
same logic and is equally straightforward to obtain~\cite{Soklakov_Forthcoming}.

The above connection between the process of learning and the associated
optimal investment growth is interesting in its own right. For the sake
of brevity we focus on just a couple of practical observations which are
important for the conclusions of this paper and leave the rest for future
discussions. The observations we have in mind can be formulated as follows.
\begin{itemize}
    \item[o1] Investor's view, $b(x)$, and the the shape of the growth-optimal
          payoff, $f(x)$, can be computed from each other (using Eqs.\/~(\ref{eq:b_from_f})
          and (\ref{eq:f_from_b}) respectively).

	\item[o2] Growth-optimal payoff $f(x)$ is a likelihood function:
	      it is proportional to the conditional probability $P({\rm research}|x)$.
\end{itemize}
Observation o1 allows us to provide the growth-optimizing
investor with a payoff structure that matches their view. More importantly,
this observation allows us to analyze any payoff function $f(x)$ by looking at
the view $b(x)$ that is associated with it. This can be useful
even if the actual investor is not growth-optimizing.
In the next section we give some examples illustrating this point and
show how observation o2 can be used for additional sanity checks.

\section*{Payoff analysis}
Before looking at example payoffs, let us outline some basic features that we
expect to see in a rational buy-and-hold investment. Imagine a
growth-optimizing investor who did no research that would challenge
the prevailing market view. Such an investor cannot justify any view
that would be materially different from $m(x)$. We must therefore have
$b(x)=m(x)$ and $f(x)=1$ for any $x$. In other words, without adequate
research any investment which is different from the risk-free $f(x)\equiv 1$
would be nothing but a pure gamble -- a gamble which cannot be justified even
by the relatively aggressive growth-optimizing investor.

Imagine now that the investor did some research
acquiring additional knowledge about the market variable
in the interval $a\leq x \leq b$. The believed distribution, $b(x)$, will now
be different from the market-implied, $m(x)$, although the two should still
agree outside the interval
\begin{equation}\label{Eq:BoundaryConditions}
\forall x\notin[a,b]\ :\ b(x)=m(x)\ \ \Rightarrow\ \ f(x)=1\,.
\end{equation}
Figure 2 illustrates a possible payoff structure satisfying this condition.

\begin{center}
    \includegraphics[scale=0.38]{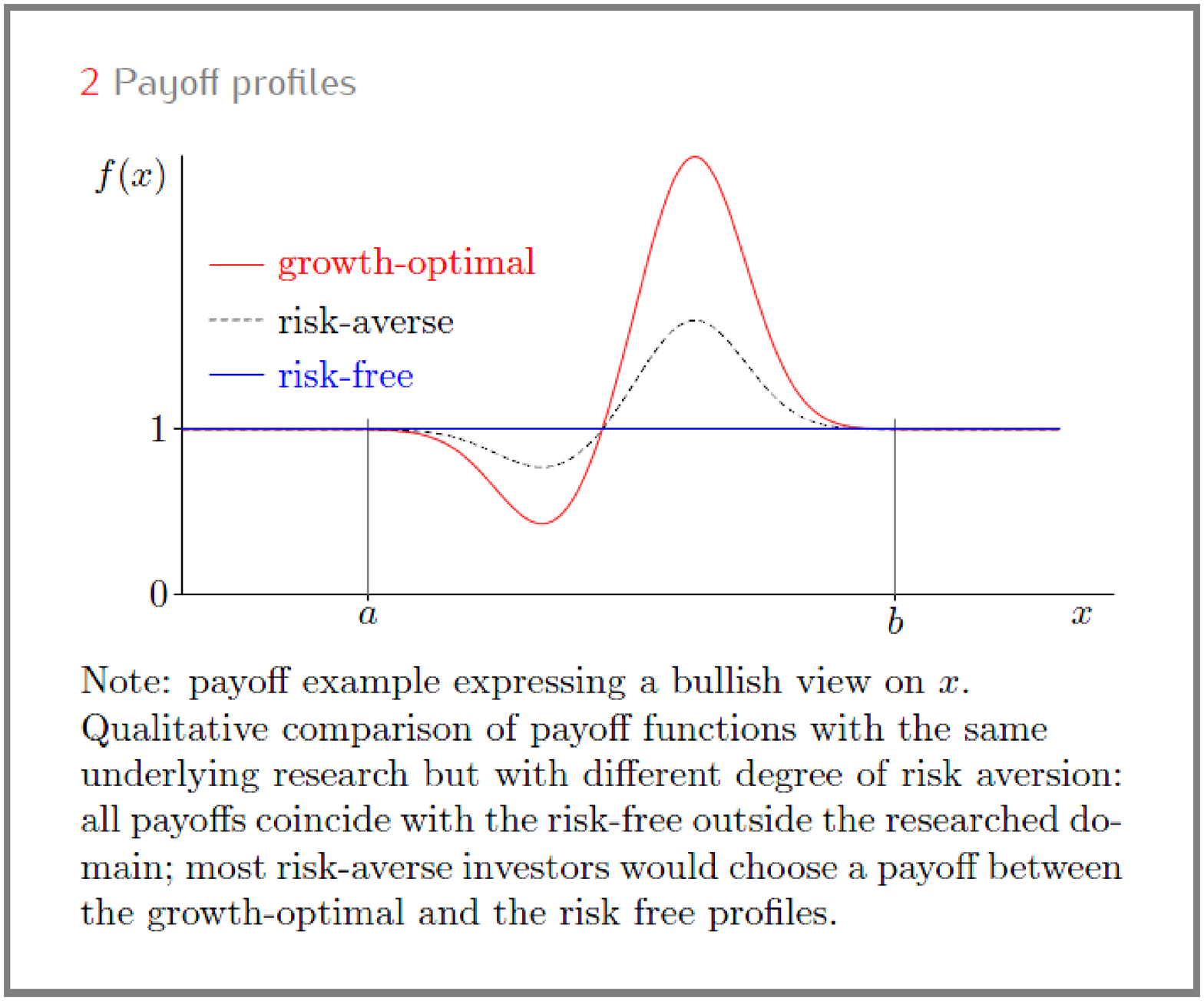}
\end{center}

How would this payoff structure change if the investor was not
growth-optimizing? As mentioned before, most investors would be more
risk averse. Achieving greater risk aversion across all values of $x$
is possible only if we chose a payoff that is closer to the risk
free line, $f(x)\equiv 1$, at every $x$.
It follows that the payoff favored by most investors
(from zero risk appetite to that of the growth-optimizing investor)
must lie between the growth-optimal and the risk-free payoffs as
illustrated on Fig.~2.

It is worth mentioning that there are limits to every research and the mathematical
extremes of market variables are all but guaranteed to fall outside these limits.
This means that some analogue of boundaries $a$ and $b$ introduced above would
always apply in practice. What can happen if we ignore this fact and extrapolate
our payoff beyond the boundaries of research? To answer this
question consider a market which moved outside the boundaries and an
investor who just realized that his positions correspond to a view that he
never researched and therefore never intended to express. Cashflows become
a matter of luck. Not many scenarios can compare to that in terms of
potential for panic, disappointment and embarrassment on all sides of the
relevant trades.

\subsection*{Simple payoffs}

Looking at Fig.~2 we cannot help noticing that very few simple products can be
tuned to comply with the boundary requirement~(\ref{Eq:BoundaryConditions}).
This is not a good sign.
Indeed, we recall that most
simple products were never really designed as investment vehicles. The original vanillas,
for example, were clearly about hedging -- ensuring businesses on future prices of goods
and supplies. Investors do however explore all available markets. Moreover, in times
of uncertainty investors naturally flee towards established hedging markets.
It is therefore prudent to include classic simple payoffs into our analysis -- especially
in the current climate when the regulatory pressure amplifies and prolongs the natural
flight to simplicity.

Figure 3 shows the relationship between market-implied and investor-believed
distributions which correspond to the at-the-money European call and digital
options. We see that investor's beliefs must be pretty far from the
market consensus to justify such products as buy-and-hold investments.
We leave the reader to judge how extreme investor-implied beliefs can become.
We note only that in the case of out-of-the-money options meaningful display of
market-implied and investor-believed distributions on the same plot becomes, typically,
impossible without introducing separate plotting scales.

\begin{center}
    \includegraphics[scale=0.38]{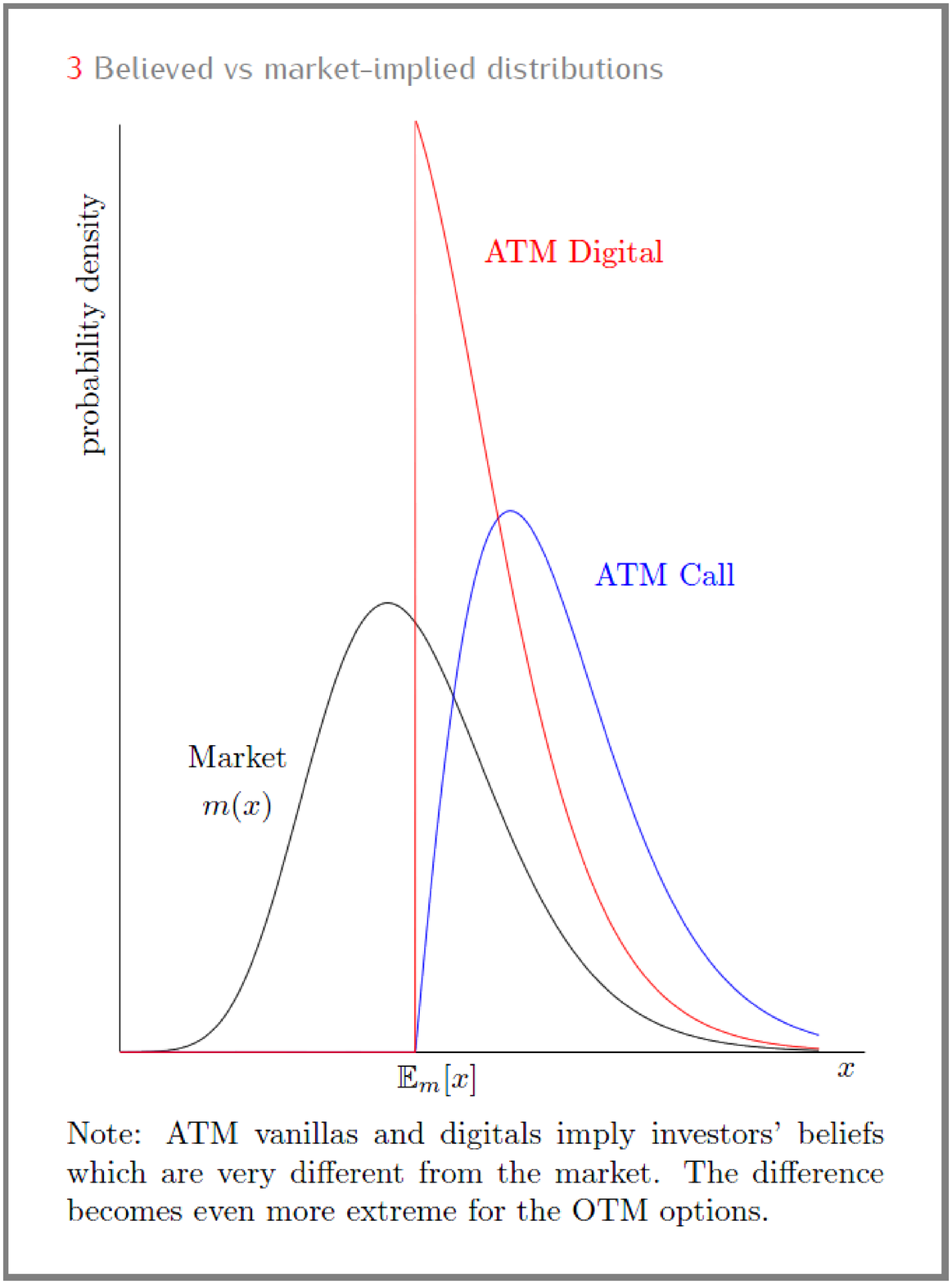}
\end{center}

Visual comparison of distributions is informative when the probabilities in question
are reasonably large. When it comes to judging rare events, however, human
intuition is notoriously poor. Looking at the tails on Fig.~3, for example,
we see that buying an at-the-money call option implies a view which is likely
to be over-leveraged on the right tail. It does not look as dramatic though as
the deviation from the market view near at-the-money. Experience tells us that
visual inspection of the tails may not give us the whole story.
Formal reasoning is required.

A more formal approach to analyzing payoff
structures can be developed from their interpretation as likelihood functions
(see observation o2 above). Note, for example, that
realistic investors cannot possibly claim that their research provides an unbounded
amount of useful information. Most of them are still confined to the decision
paradigm described by just a few possible outcomes
(e.g. \lq\lq buy", \lq\lq hold" or \lq\lq sell"). In mathematical terms this means that \lq\lq research"
in Eq.~(\ref{Eq:Bayes}) belongs to a discrete finite set of outcomes.
This implies that $P({\rm research}|x)$ is bounded and $f(x)$ cannot diverge
to infinity. We conclude that anyone with an initially neutral position who uses a call option as
a buy-and-hold investment is behaving in an
economically indistinguishable way from a growth-optimizing investor with views
so extreme that no research could be found to back them up (without
contradicting the basic laws of probabilities).

Despite all we have said above, vanilla products are perceived by investors as a
relatively safe choice among derivatives. This is because of the huge hedging
market which is formed by these products. Liquidity provided by the market
allows investors to dump their investments without having to wait until maturity.
Liquidity is a good thing to have, but excessive substitution of
liquidity in place of prudent investment design is dangerous.
We have seen that the use of vanillas as investment vehicles implies
rather extreme views. Widespread expression of such views signals expectation of
instability. This expectation must get reflected by the market: after all, we are
talking about real investors placing real money on extreme market events.
In the limiting case we could end up with a crash, but in any case the market
should price investors' behavior and we should see fattening of the tails in the
return distributions.\footnote{Could this be what happened in the crash of 1987?
Black-Scholes inspired technologies lead to unprecedented growth in speculative
trading of vanilla products. Market crashed, but quickly stabilized around
a new type of equilibrium with considerably fatter tails (volatility skew).}

\section*{Information derivatives}
In this section we touch a bit more upon the question of safety. An honest investor
should be prepared to make some losses (in real terms) if their view turns
out to be less accurate than the market. A lot of statistical trading strategies,
however, can lead to losses even if the underlying view is exactly right.
You may, for example, hold the perfectly accurate belief that an unbiased coin
flip has 1/2 chance of producing heads or tails and still struggle to stay in the
game through a particularly unlucky sequence of coin flips. In this section we present
an idea for how this kind of risk can be reduced with the help of derivative technologies.

It can be shown that the expected performance of the growth-optimizing strategy
can be achieved in the long run by simple repetition: by reinvesting proceeds
from one investment into the next~\cite{Kelly_1956}. This, of course,
is rather risky in practice. Instead of
performing such reinvestments for real, we propose to compute the expected
performance mathematically and use the resulting formula as a definition
of a derivative product. Effectively, we propose to skip the bulk
of risky and expensive reinvestments and, instead of waiting for the
law of large numbers to kick in, we offer direct exposure to the
expected payoff. Statistical (i.e. risky) convergence on the investor's side
is thereby replaced with derivative hedging on the sell side. This provides
added value to the investor and uses much safer technology (hedging vs.statistical convergence).

The easiest way to explain this idea is by giving an example.
In the following we deliberately bring up a few important
points from the previous sections to show how they all fit together.
Consider a market which believes in a lognormal distribution for variable $x$
\begin{equation}
   m(x)={\rm LN}(\mu,\sigma_m)\stackrel{\rm def}{=}\frac{1}{x\sqrt{2\pi\sigma_m^2}}\exp\Big[-\frac{(\ln x -\mu)^2}{2\sigma_m^2}\Big]\,.
\end{equation}
Imagine now an investor who agrees with
the market on everything apart from the variance which he believes
to be underestimated by the market. This kind of view relative to the market is very
plausible (if not common). Mathematically, the investor-believed distribution must be
$b(x)={\rm LN}(\mu,\sigma_b)$ where $\sigma_b>\sigma_m$. An attentive reader would
immediately spot a problem: the growth-optimal payoff
\begin{equation}\label{Eq:Variance_f}
   f(x)=\frac{b(x)}{m(x)}=\frac{\sigma_m}{\sigma_b}\exp\Big[(\frac{1}{\sigma_m^2}-\frac{1}{\sigma_b^2})\frac{(\ln x-\mu)^2}{2}\Big]
\end{equation}
is unbounded in the wings contradicting Eq.~(\ref{Eq:BoundaryConditions}).
By observation o2, even the growth-optimizing investor would have to \lq\lq bend the rules" of conditional
probabilities to justify this kind of leverage in the wings. Let us proceed with this view anyway
and see where it would lead us. This simple example gives us the opportunity to trace the propagation
of over-leveraging in the wings from a seemingly innocent belief to the final product.
We then discuss how this over-leveraging can be avoided or mended.

Substituting (\ref{Eq:Variance_f}) into (\ref{Eq:ERate}) and introducing the concept
of realized variance $\sigma_p^2$ by writing $p(x)={\rm LN}(\mu,\sigma_p)$ we obtain
\begin{equation}\label{Eq:VarSwap}
\set{E}_p[{\rm rate}] = \int_0^\infty p(x)\ln\frac{b(x)}{m(x)}\,dx \propto \sigma_p^2 + {\rm const}\,,
\end{equation}
which is the payoff structure of a variance swap.
With $\sigma_b>\sigma_m$, the coefficient before $\sigma_p^2$ is positive, suggesting the investor takes a long position in the swap.

Theoretically, variance swaps are among the best understood products.
In practice, however, traders are all too familiar with difficulties that come
with managing this product~\cite{ChrissMorokoff_1999}. Although the above investment
scenario is simplistic, our approach succeeds in raising a warning which turns out to be
true in a much wider context. Moreover, it gives us the following high-level explanation
of the problem. Variance is a global characteristic of a distribution. Consequently, variance swaps
express a view across the whole of the distribution including the wings. This includes
extreme values of $x$ which fall outside any finite range covered by research.
On the hedging side this looks like
exposure to all strikes from zero to infinity -- including the strikes which
fall outside the finite range available for practical hedging. No amount of traditional modeling is going to change
this fact and the solution to such problems is clearly in the domain
of product design.

Indeed, there is now a variety of modifications to the basic
variance swap which attempt to address the shortcomings of the original
design (see, e.g.~\cite{CarrLewis_2004}). Capped, conditional, corridor
and gamma swaps have been introduced. Index variance swaps have been the longest
used example of the original payoff structure. They too have recently developed
caps on the account of hedging problems that became apparent during the last crisis.
For single-stock variance swaps such caps proved to be insufficient and
the market is once again exploring alternative payoff structures~\cite{Clark_2010}.
The market is clearly looking for prudently designed investment structures but,
to our shame, realized losses are still setting the trends of this development.

As the above discussion shows, we must get into the habit of designing products
which acknowledge the limits of available information. This can be done either
explicitly, by checking the asymptotic behavior (\ref{Eq:BoundaryConditions}),
or via payoff-related modeling, as considered in~\cite{Soklakov_2008}.
These two approaches are of course related: they both rely on concepts of information
and information processing and even lead to some similar product designs.
The reason for this is very simple: for investment purposes, information
is the ultimate asset class. Information production, processing and packaging into
financial products are just natural facets that we must examine when building,
using and regulating the use of investment products.

\section*{Summary and outlook}

We presented a framework for creating new and analyzing existing
buy-and-hold investment structures. We did that using very simple arguments and
hardly any mathematics. As a result, all product complexity becomes directly
linked to that of investor's research, market data and, if you want to use
the technique of the last section, the general form of a future realized
distribution. No exotic features would appear in the final product
unless they are evident from investor's research or market data.
Barriers and digitals, for example, would not appear
unless investor's beliefs or market data were best described by
discontinuous probability distributions. In spite of this push to
simplicity, we find that very simple payoffs can struggle to
match realistic beliefs of a buy-and-hold investor. Judging by~\cite{Soklakov_2008},
the complexity of basic information derivatives is similar to that of
third generation volatility products (e.g. gamma swaps).

On the technical
side we show that optimal payoff structures are intimately connected to
the central concept in statistical inference, -- the likelihood functions.
This can be useful for transforming results of statistical research
into tradable payoff structures. In terms of tools for further research,
information theory, learning theory and utility framework should provide the
core technical concepts.

In this paper we steered as far away from modeling as we could. This was done
to emphasize the importance of model-independent aspects of product design.
Models, however, do influence the design of investment products.
Our framework is well positioned for analyzing such influence.
More generally, our approach can be applied to rationalizing the use of
different models. Indeed, a calibrated model is nothing but a sophisticated
definition of a
probability distribution. Models can be used to express a view on the market
just like the probability distributions that we considered above.
We can expose model differences by designing products that exploit them.
This is of course likely to require more advanced analysis, but only as
advanced as the models in question. Finally, our approach may add motivation
and ideas to modeling approaches based on information processes~(see e.g.~\cite{BrodyFriedman_2009}).

$\blacksquare$

\vspace*{2mm}

{\small
Andrei Soklakov is a Vice President in Model Risk and Analytics of Deutsche Bank.
He would like to thank his former and present colleagues at both Deutsche Bank and
Goldman Sachs for their interest in this work. The core setup underlying the results
of this paper was first presented by the author at Quant Congress Europe 2008. The views
expressed herein should not be considered as investment advice or promotion.
They represent personal research of the author and do not necessarily reflect
the view of his employers, or their associates or affiliates. Email: Andrei.Soklakov@db.com, Andrei.Soklakov@gmail.com.}

\section*{Appendix}

Equation~(\ref{eq:b_from_f}) and its interpretation in terms of Bayesian learning
is probably the most important logical element of this paper. We arrived at it by
considering a growth-optimizing investor and by making two further assumptions: all
of the capital is invested and $\sum_i1/R_i=1$. It turns out that none of these
assumptions are necessary. Somewhat amazingly, the same core logic holds even
in the case of an arbitrary rational investor. The case of general investors requires
a separate presentation and we will provide it in a follow-up
paper~\cite{Soklakov_Forthcoming}. Here we give a brief summary of what happens
if we remove all additional assumptions in the case of growth-optimizing investor.

If all of the investment capital is committed, $\sum_i\alpha_i=1$, then the
optimization problem, Eqs.(\ref{Eq:EbRate},\ref{Eq:Alpha*}), does
not depend on whether $\{1/R_i\}$ add up to one or not. The only thing that
needs fixing is the normalization of market-implied probability. To this
end we compute the reference return, $R$, by setting $1/R=\sum_i1/R_i$
and replace the previous definition of the market-implied probability with
a more general one: $m_i\stackrel{\rm def}{=}R/R_i$. All arguments follow as before
and Eq.~(\ref{eq:b_from_f}) becomes
\begin{equation}\label{eq:b_from_f2}
b(x)=\frac{f(x)}{R}\,m(x)\,.
\end{equation}
This leaves intact the interpretation of the payoff, $f(x)$, as the likelihood function
(although, of course, one should be careful with numerical details: payoff asymptotics, etc.).

We now remove the remaining assumption and allow the investor to keep some of the investment
capital risk-free (growing at the rate $R_0$). This becomes relevant when $R<R_0$, i.e. when the
risky returns offered by the market appear less attractive than risk-free. Mathematically, this case
is very similar to the case of \lq\lq unfair odds''~(see Ref.~\cite{Kelly_1956}) -- we just need to introduce
the concept of risk-free rate, $R_0$, which is not present in the classic betting paradigm.
As expected, the investment becomes a sum of risky investments on some values of $x$
(lets denote them $\omega$) and a risk-free floor. We derive
\begin{equation}\label{Eq:General}
b(x)=\frac{f(x)}{R_0}\,m^*(x)\,,{\rm\ \ \ and \ \ }
m^*(x)\stackrel{\rm def}{=}\left\{\begin{array}{ll}
                R_0/R_{i(x)}\,,&x\in\omega,\cr
                b(x)/\alpha_0\,,&x\not\in\omega,
              \end{array}
        \right.
\end{equation}
where $m^*(x)$ is a probability distribution and the normalization constant $\alpha_0$ coincides with the fraction
of the capital invested in a risk-free bond. The case $x\in\omega$ is as clear as Eq.~(\ref{eq:b_from_f2}).
To understand what happens for $x\not\in\omega$, imagine coming to a fruit market and noticing an ordinary
apple being on sale for one billion dollars. You cannot accept that price even as prior information for
your research.
Similarly, investment market for $x\not\in\omega$ does not really exist -- the risky returns are far to low
(as compared to risk-free). The market simply does not offer any reasonable prior view and the investor patches it
with the only view available -- his own belief. In terms of the payoff for $x\not\in\omega$, $b(x)$ drops out of
Eq.~(\ref{Eq:General}) leaving the equation for the risk-free payoff $f(x)=\alpha_0R_0$ (just like it should
be in all cases when the investor's research cannot proceed). We see that our framework makes sense even
when investment markets struggle to exist.

\end{document}